\documentclass[onecolumn,useAMS,usenatbib]{mn2e}
\usepackage{graphicx}
\usepackage{amsmath}
\usepackage{amssymb}
\title[A kinematical approach to gravitational lensing]
  {A kinematical approach to gravitational lensing using new formulae for refractive index and acceleration}
\author[Walters, Forbes \& Jarvis]
  {S.J. Walters, L.K. Forbes and P.D. Jarvis \\
  School of Mathematics and Physics, University of Tasmania, P.O. Box 37, Hobart, 7001, Tasmania, Australia}
\date{Released 2010 Xxxxx XX}

\begin{document}

\maketitle

\begin{abstract}
This paper uses the Schwarzschild metric to derive an effective refractive index and acceleration vector that account for relativistic deflection of light rays, in an otherwise classical kinematic framework. The new refractive index and the known path equation are integrated to give accurate results for travel time and deflection angle, respectively. A new formula for coordinate acceleration is derived which describes the path of a massless test particle in the vicinity of a spherically symmetric mass density distribution. A standard ray-shooting technique is used to compare the deflection angle and time delay predicted by this new formula with the previously calculated values, and with standard first order approximations. Finally, the ray shooting method is used in theoretical examples of strong and weak lensing, reproducing known observer-plane caustic patterns for multiple masses.
\end{abstract}

\begin{keywords}
 gravitation -- gravitational lensing: micro -- methods: numerical -- acceleration of particles -- planets and satellites: detection -- gravitational lensing: strong.
\end{keywords}

\section{Introduction}

The deflection of light by massive bodies has been of interest to mathematicians and physicists from time to time since Newton suggested the possibility (\citet{new}). This deflection was calculated in the late eighteenth and early nineteenth centuries, treating light as a classical particle. The deflection was again calculated by Einstein in the early twentieth century, using his new general theory of relativity, to be twice the previous classical result. The measurement of the deflection of light passing close by the sun was widely publicized as a dramatic confirmation of general relativity, in the now famous 1919 expeditions.

In the last three decades, gravitational lensing has become an important tool for astrophysicists, especially in searching for dark matter (first suggested in \citet{pac}) and exoplanets. By 1991, astrophysicists were suggesting that exoplanets could be found using microlensing, and since 2004 at least ten planets have been
found in this way (\citet{sumi}). In microlensing, light from a background star passes close to the lensing system, and is deflected around the lens. Because of this, more light rays reach the observer, producing magnification of the background source. This magnification changes over time, as the source, lens and observer move into and out of alignment. The details of the magnification over time are plotted in a `light curve', which is simply intensity versus time. 
A planet orbiting a lensing star can make changes in the magnification
pattern at the observer's plane. Such changes show up as variations to the shape of the simple light curve. 

Interpretation of these light curves is difficult, as this is an inverse problem, in which observers seek to determine the details of the lensing system which gave rise to the observed data.
In particular, researchers are often trying to find planets in the lensing system, and to determine characteristics of such planets, primarily orbital radius and mass. Details on reproducing a model
of such a multi-body lens were outlined at least as early as 1996 (\citet{wam}). For a brief history see \citet{sch}, pages 1-2, \citet{mao}, or the recent review by \citet{ell}. In this paper, microlensing will be considered, although the formulae derived here are just as valid for macrolensing situations. 

\subsection{Lensing Models}

It is customary (\citet{wam}) to use a `thin lens' model, in which the effect of the lensing system is confined to the plane containing the lensing objects (the `lens plane'), this plane being normal to the line from observer to lens (or alternatively, source to lens). A deflected light ray thus consists approximately of two straight lines, with an abrupt angle change in the lens plane. The magnitude of this change is given by a simple formula: $\Delta\theta=2r_{s}/r_{0}$ where $r_{0}$ is the point of closest approach of the light ray to the star or planet and $r_{s}$ is the Schwarzschild radius, $r_{s}=2MG/c^{2}$. Here, $M$ is the mass of the star, $G$ is Newton's gravitational constant and $c$ is the speed of light in a vacuum.

\begin{figure}
\vspace{1cm}
\caption{Thin Lens Approach and Typical Light Curve}
\includegraphics[height=5cm]{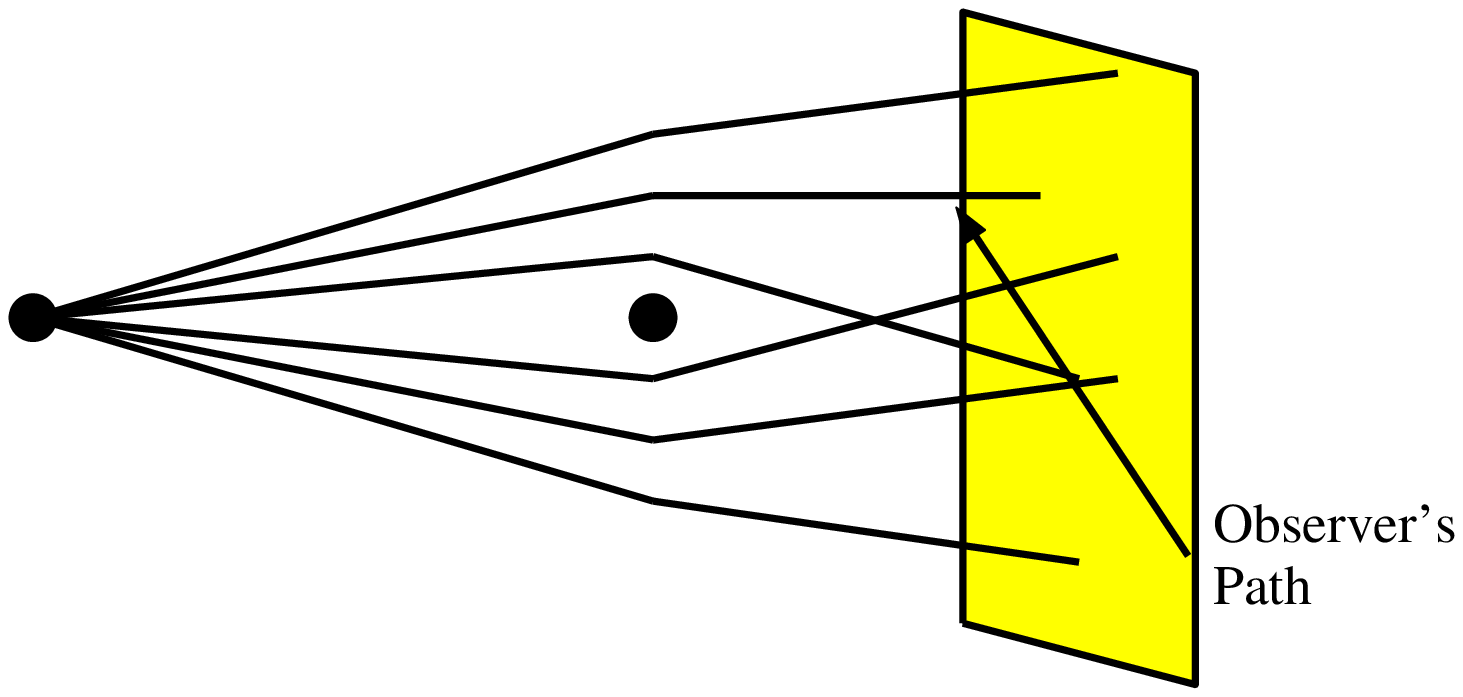}
\quad
\quad
\includegraphics[height=5cm]{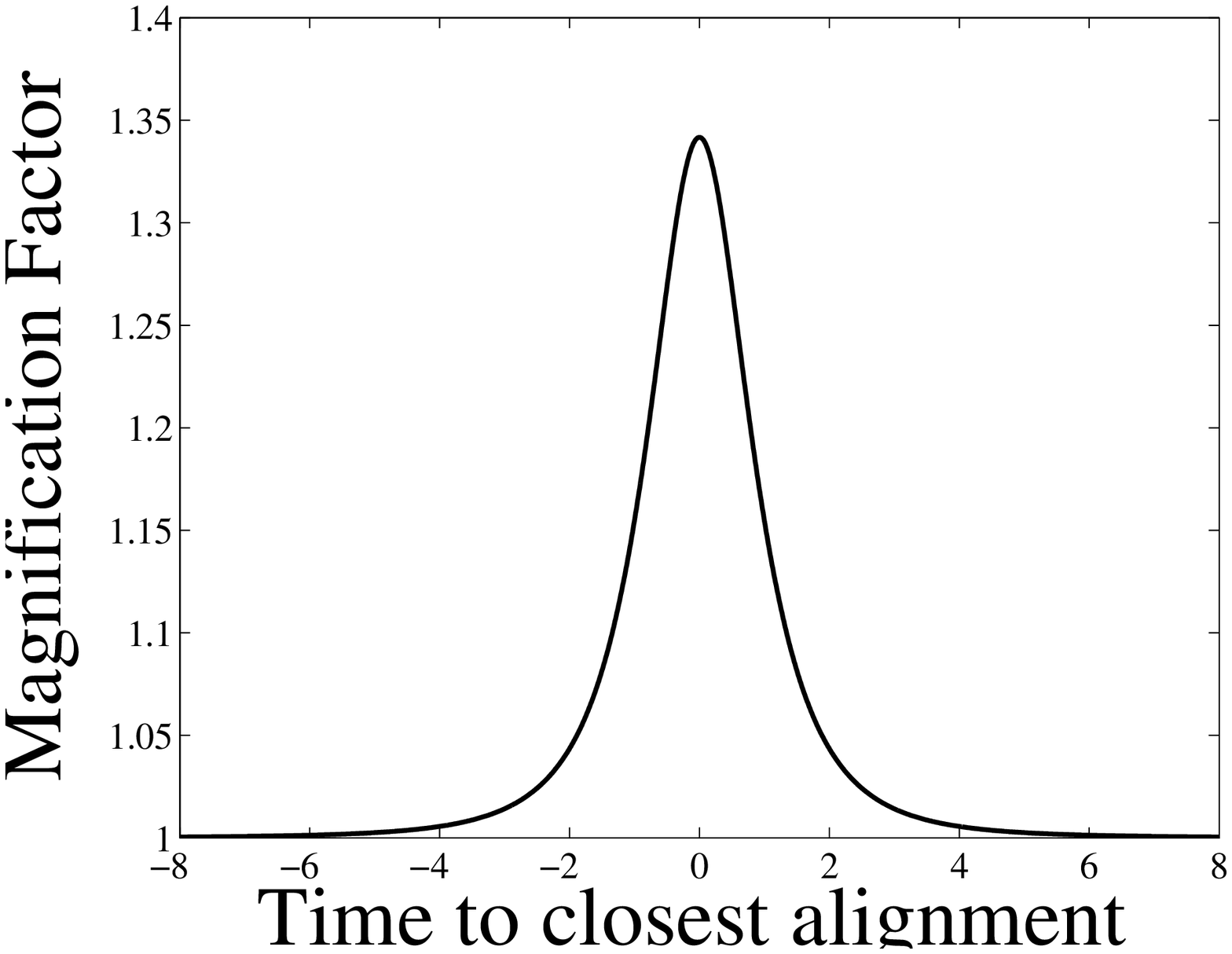}
\label{curves}
\end{figure}

Current methods (\citet{zab} and \citet{wam}) shoot rays from the observer to the lens plane, deflect by the angle as calculated above, and then draw the ray from there to the source plane. Equivalently, light rays may be followed from the source to the observer, mimicking more closely the actual physics. The density of rays at the source plane (alternatively, the observer's plane) is thus mapped. By tracing various linear paths across this `magnification map', to simulate the relative movement of the source star, light curves are generated by plotting density of rays as a function of time. Fig. (1) shows schematically the method for mapping the amplification for an observer travelling at constant speed in the observer's plane. Simple light curves such as the one in the right of Fig. (\ref{curves}) provide information about the mass of the lensing object, provided that the distance to lens and distance
to source can be estimated. If the lens is a binary system, there are deviations from this simple light curve, which may yield information about the orbital distance and relative masses in the system.

Thus the purpose of a microlensing model is to produce a magnification map due to light being deflected by the lensing system. Note that every such
magnification map allows for arbitrarily many light curves, depending on the location of the observer's (or source's) path across the plane. Because of this, it is, in general,
a difficult matter to find a model to fit an observed light curve
for a lens involving more than one mass (that is, a binary or planetary
system). The problem becomes much more difficult when the lens involves
more than two objects (\citet{mao}).

The approach presented in this paper is to generate a magnification map by shooting rays through a lens of smoothly
varying refractive index. The path will be curved rather than two straight line segments. Even so, the majority of the
deflection occurs very close to the lens plane.

\subsection{Kinematical Approach}

It is the purpose of this paper to consider gravitational lensing using an alternative approach. This approach can be described as an equivalent classical kinematical formulation, which nevertheless replaces Newton's formula for gravitational acceleration with a formula derived from general relativity. Using this kinematical approach, each curved light ray is traced from the source to the plane of the observer, rather than using the simple angle change formula commonly used in lensing models.

We will start by considering the path predicted by general relativity for a massless test particle (a photon) near a massive body. The changes in angle along this path will be integrated to give the deflection of the light ray predicted by general relativity. As a check, we will compare this result with known formula for approximating the deflection to first order in the relevant small constant. Next we will derive an `effective refractive index' due to the massive body. Integration will be used to predict the Shapiro delay, which will also be checked against the known first order prediction of general relativity. A new acceleration formula will then be derived and tested against these known values for the deflection and delay using a forward integration method. Finally, we will describe and exemplify a procedure for using the new acceleration formula to produce magnification maps for multi-body lenses.

\section{Light Paths in a Schwarzschild System}

The Schwarzschild metric describes spacetime outside a non-rotating, electrically neutral, spherically symmetric mass density distribution $M$ and is taken as a valid approximation for local spacetime structure in the vicinity of any massive body (stellar systems, including black holes, but also planets) having negligible charge and angular momentum. In spherical coordinates the metric components are given in terms of the invariant interval $dl^{2}$ as (\citet{mtw})
\begin{equation}
dl^{2}=\frac{r-r_{s}}{r}c^{2}dt^{2}-\frac{r}{r-r_{s}}dr^{2}-r^{2}(\sin^{2}\theta d\phi^{2}+d\theta^{2}),
\label{sm}
\end{equation}
where $r_{s}=2MG/c^{2}$ is the Schwarzschild radius, $G$ is Newton's constant, and $c$ is the constant speed of light in vacuum. Here $r$ , $\theta$,  $\phi$ are   coordinates which correspond to standard spherical coordinates in the reference frame of an observer at rest far from the system.

Light travels on null geodesics with $dl=0$, so equation (\ref{sm}) becomes
\begin{equation}
cdt=\frac{r}{r-r_{s}}\sqrt{dr^{2}+r(r-r_{s})(\sin^{2}\theta d\phi^{2}+d\theta^{2})}.
\label{cdt}
\end{equation}
Applying Fermat's principle, that light follows a path that extremizes travel time $T$, we can consider the functional:
\begin{equation}
T=\int dt=\frac{1}{c}\int\frac{r}{r-r_{s}}\sqrt{dr^{2}+r(r-r_{s})(\sin^{2}\theta d\phi^{2}+d\theta^{2})}.
\label{fermat}
\end{equation}

To illustrate the use of equation (\ref{fermat}) we firstly derive a path equation which is equivalent to the usual trajectory equations for null geodesics, and numerically integrate it to give the deflection of a light ray according to general relativity.

Without loss of generality, for a single light ray, the coordinates can be oriented so that the ray is in the plane $\theta=\pi/2$.
Then $d\theta=0$ in equation (\ref{fermat}), which consequently may be re-arranged to give

\begin{equation}
T=\frac{1}{c}\int\frac{r}{r-r_{s}}\sqrt{1+r(r-r_{s})\phi'^{2}}dr
\label{time2}
\end{equation}
adopting the notation $\phi'=d\phi/dr$ . Let $F$ be
the integrand in equation (\ref{time2}), that is: 
\begin{equation}
F(r,\phi,\phi')=\frac{r}{r-r_{s}}\sqrt{1+r(r-r_{s})\phi'^{2}}.
\label{int}
\end{equation}
Then the Euler-Lagrange equation is \begin{equation}
\frac{\partial F}{\partial\phi}-\frac{d}{dr}(\frac{\partial F}{\partial\phi'})=0.\label{eq:11}\end{equation}
It can be seen in equation (\ref{int}), that there is no explicit
dependence of $F$ upon $\phi$ so clearly $\partial F/\partial\phi=0$.
Therefore this term vanishes in equation (\ref{eq:11}) which then has
the immediate first integral
\[
\frac{\partial F}{\partial\phi'}=\frac{r^{2}\phi'}{\sqrt{1+r(r-r_{s})\phi'^{2}}}=K,\]
in which $K$ is a constant. Now $1/\phi'=dr/d\phi$, so rearrangement yields the first order separable ODE
\[
\frac{dr}{d\phi}=\pm\sqrt{\frac{r^{4}}{K^{2}}-r(r-r_{s})}.
\]
This constant $K$ can be determined. At the point of closest approach to
the mass (call this point $r=r_{0}$), the radius is at a minimum,
that is, $dr/d\phi=0$, so it follows that: 
\[
K^{2}=\frac{r_{0}^{3}}{r_{0}-r_{s}}.
\]
Thus the path is defined by:
\begin{equation}
\frac{dr}{d\phi}=\pm\sqrt{\frac{r^{4}(r_{0}-r_{s})}{r_{0}^{3}}-r(r-r_{s})}.
\label{eq:drdphi}
\end{equation}
It can be easily seen that substituting $u=1/r$ followed by differentiation gives the well known second order equation as (\citet{cap})

\begin{equation}
\frac{d^2u}{d\phi^2}+u=\frac{3r_{s}}{2}u^2.
\end{equation}

From equation (\ref{eq:drdphi}), an integral can easily be written to evaluate
the total change in $\phi$ of a light ray passing a massive object, which will
be twice the change from the perihelion out to infinity:
\[
\triangle\phi=2\intop_{r_{0}}^{\infty}\frac{dr}{\sqrt{r^{4}(r_{0}-r_{s})/r_{0}^{3}-r(r-r_{s})}},
\]
This elliptic integral cannot be evaluated with a finite number of
simple algebraic terms. Numerical integration could be used to evaluate
the deflection. However, the integrand is infinite at $r=r_{0}$,
making the accuracy of any numerical evaluation questionable. By using
a substitution $r=1/\cos (2\psi)$ , the singularity can be removed.
After simplification we obtain:
\[
\triangle\phi=4\sqrt{\frac{2r_{0}}{r_{s}}}\intop_{0}^{\pi/4}\frac{d\psi}{\sqrt{2r_{0}/r_{s}+2-4\cos ^{2}\psi-\sec^{2}\psi}}.
\]
This integrand is well behaved over the interval, so it can be numerically
integrated to any desired accuracy, for example, by gaussian quadrature.
An undeflected ray will have $\triangle\phi=\pi$, so the deflection for a
ray will be $\delta=\triangle\phi-\pi$. The deflection is usually very small, so in order to avoid a {}`loss of significance error', this subtraction should be performed in the integrand. For readability, let $\alpha=2r_{0}/r_{s}$, and let $\beta=2-4\cos^{2}\psi-\sec^{2}\psi$.
In this notation, the deflection $\delta$ can therefore be expressed as
\[
\delta=4\sqrt{\alpha}\intop_{0}^{\pi/4}(\frac{1}{\sqrt{\alpha+\beta}}-\frac{1}{\sqrt{\alpha}})d\psi.
\]
The two terms in the integrand are combined, so that the subtraction can be performed explicitly. This yields:
\begin{equation}
\delta=4\intop_{0}^{\pi/4}(\frac{-\beta}{\sqrt{\alpha+\beta}(\sqrt{\alpha}+\sqrt{\alpha+\beta})})d\psi.
\label{deflect1}
\end{equation}
For the purposes of this paper, we take the solar radius to be 696000 kilometres, and the solar Schwarzschild radius to be 2.95 kilometres. Numerical integration of equation (\ref{deflect1}) for a ray passing near the surface of the sun ($r_{s}/r_{0}=2.95/696000$))gives a deflection of $1.74851634161261$ arcseconds. As the path equation contains all of the information about the general relativistic path of the photon, this is the deflection predicted by general relativity to the level of precision shown. This deflection angle will be used later to confirm the accuracy of the kinematic approach.
As a check, we can consider Einstein's estimate for the deflection (\citet{sch}, page 3)
\begin{equation}
\frac{2r_{s}}{r_{0}}=1.74850913341648 \textnormal{ arcseconds}.
\label{deflect2}
\end{equation}
This estimate is found to correspond to our calculated value to first order in $r_{s}/r_{0}$, as expected.

\section{New Refractive Index and Travel Time Delay}

The approach presented here makes use of an expression for the
refractive index, $n$. The Schwarzschild metric from section
2 will be used to derive an `effective' refractive index.
As suggested above, the functional (\ref{fermat}) can be arranged in the form $T=\int dt=\int(dt/ds)ds$, where $s$ is an arbitrary parametrization of the ray path. By choosing $ds$ to be the element of arc-length along the path, the speed is then $v=ds/dt$, and thus the refractive index is $n=c/v=c dt/ds$, and we obtain finally $T=1/c \int nds$, with the refractive index then having the form
\[
n=\frac{r}{r-r_{s}}\sqrt{r'^{2}+r(r-r_{s})(\sin^{2}\theta\phi'^{2}+\theta'^{2})},
\]
where $r'=dr/ds$, $\theta'=d\theta/ds$ and $\phi'=d\phi/ds$. (Note that there is no suggestion here of any physical effect on the speed of light - indeed, any local measurement of coordinate velocity is guaranteed to result in the usual speed $c$).

Taking again the two dimensional case, with $\theta=\pi/2$,
the refractive index has the form
\begin{equation}
n=\frac{r}{r-r_{s}}\sqrt{r'^{2}+r(r-r_{s})\phi'^{2}}.
\label{eq:ref2d}
\end{equation}

\begin{figure}
\vspace{1cm}
\caption{photon path near the sun}
\includegraphics[width=10cm]{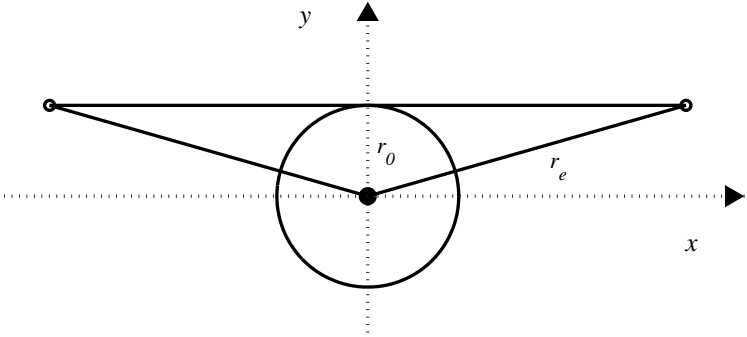}
\label{delay}
\end{figure}

As an example, the delay can be calculated for light
to travel from an object at earth radius, skim past the sun, and back out
to earth radius. This is the calculation needed for the radar echo delay test of general relativity (\citet{shapiro}). (In fact, for that particular test, it would be necessary to consider another planet such as Venus or Mars, and the calculation would need to be performed for each leg of the journey. For the purposes of this paper, it is simpler to imagine a reflecting satellite at the same orbit as the earth). The problem can be pictured as in Fig. \ref{delay}, not
to scale. The deflection is
small, so that the path appears as a straight line. In fact the curved path that the photon takes is derived above in equation (\ref{eq:drdphi}).
The path is symmetric about the point of closest approach ($r_{0}$),
so that exactly half the delay can be obtained by integrating from $r_{0}$ to $r_{e}$, the radius of the earth's
orbit. If the sun had no effect on the light path, the distance
from perihelion to earth would be the straight line distance $\sqrt{r_{e}^{2}-r_{0}^{2}}$.

The delay can be calculated using the new refractive index, equation (\ref{eq:ref2d}), and multiplying by $ds$ so that 
\[
nds=\frac{r}{r-r_{s}}\sqrt{dr^{2}+r(r-r_{s})d\phi^{2}}=\frac{r}{r-r_{s}}\sqrt{1+r(r-r_{s})(\frac{d\phi}{dr})^{2}}dr.
\]
Again, $(d\phi/dr)^{2}$ is given by the path equation (\ref{eq:drdphi}).
The time taken for the trip will be twice the time to go from the
point of closest approach to the sun, $r_{0}$ to the earth's orbit,
$r_{e}$, that is
\[
T=\frac{2}{c}\int_{r_{0}}^{r_{e}}nds=\frac{2}{c}\int_{r_{0}}^{r_{e}}\frac{r}{r-r_{s}}\sqrt{1+\frac{r(r-r_{s})}{r^{4}(r_{0}-r_{s})/r_{0}^{3}-r(r-r_{s})}}dr.
\]
The integrand is infinite at $r=r_{0}$, so before integrating, we make the substitution $\cos(\psi)=r_{0}/r$. After rearrangement we get:

\begin{equation}
T=\frac{2r_{0}^2}{c}\int_{0}^{\arccos(r_{0}/r_{e})}\frac{d\psi}{\cos^{2}\psi(r_{0}-r_{s}\cos\psi)}\sqrt{\frac{(r_{0}-r_{s})(1+\cos\psi)}{(r_{0}-r_{s})(1+\cos\psi)-r_{s}\cos^{2}\psi}}.
\label{delay1}
\end{equation}
This integrand is perfectly well behaved over the interval, so it
can be integrated to arbitrary precision by using, for example, Gaussian
Quadrature. As for the deflection, the delay is very small, so it is important to subtract the straight-line time before
integrating. The straight-line time can be written as:

\begin{equation}
T_{0}=\frac{2}{c}\sqrt{r_{e}^{2}-r_{0}^{2}}=\frac{2r_{0}}{c}\int_{0}^{\arccos(r_{0}/r_{e})}\frac{d\psi}{\cos^{2}\psi}.
\label{delay2}
\end{equation}
The delay is obtained by subtracting equation(\ref{delay2}) from equation(\ref{delay1}) to give:

\begin{equation}
\triangle t=\frac{2r_{0}}{c}\int_{0}^{\arccos(r_{0}/r_{e})}\frac{(1-\gamma)d\psi}{\gamma\cos^{2}\psi}
\label{delay3}
\end{equation}
where 
\[
\gamma=\bigl(1-(r_{s}/r_{0})\cos\psi\bigr)\sqrt{1-\frac{r_{s}\cos^{2}\psi}{(r_{0}-r_{s})(1+\cos\psi)}}
\]
This expression (\ref{delay3}) is now in a form that minimizes errors caused by subtraction of large terms.
Performing this calculation in Matlab gives a delay of $129.0896086$ $\mu s$.
As for the deflection, this calculation can be performed to arbitrary
precision, and accurately describes the delay predicted by general relativity.

For comparison, \citet{wei}, pages 201 - 203, gives the following formula
for the delay, to first order in $r_{s}/r_0$:
\begin{equation}
\triangle t=\frac{r_{s}}{c}\biggl[2\ln\biggl(\frac{r_{e}+\sqrt{r_{e}^{2}+r_{0}^{2}}}{r_{0}}\biggr)+\sqrt{\frac{r_{e}-r_{0}}{r_{e}+r_{0}}}\biggr].
\label{eq:weinberg1}
\end{equation}
This gives a delay of $129.0894053$ $\mu s$. The fractional
difference between this and the delay calculated above using equation (\ref{delay3}) is:
\[
\frac{129.0896086-129.0894053}{129.0896086}=1.57\text{x}10^{-6},
\]
which is of the same order as $r_{s}/r_{0}=2.95/696000\simeq4.24\text{x}10^{-6}$, as expected. Note that
equation (\ref{eq:weinberg1}) is usually further approximated and simplified
(\citet{wei}, page 203, \citet{ken}, pages 95-96) by saying that
$r_{0}<<r_{e}$. This gives:
\[
\triangle t=\frac{r_{s}}{c}\biggl[2\ln\biggl(\frac{2r_{e}}{r_{0}}\biggr)+1\biggr],
\]
which is significantly less accurate, giving a delay of $129.1350325$ $\mu s$.

We now turn to a kinematic approach, and consider differential equations that relate position, velocity and acceleration. In a Newtonian system, the acceleration would be given by Newton's law of gravitational attraction, $g=-c^2r_{s}/2r^{2} \boldsymbol{e_{r}}$. This simple formula is not appropriate for the present application, and a new form will now be derived from the present relativistic approach.

\section{A New Acceleration Formula}

By combining the metric equation (\ref{sm}) with the path equation (\ref{eq:drdphi}) for a photon in a Schwarzschild orbit, the
velocity and acceleration of the photon due to the nearby mass are now derived. Here, the meaning is that of a coordinate acceleration. As a freely falling particle, the photon does not experience any locally measurable force. 
Beginning with the path equation (\ref{eq:drdphi}), and making the substitutions

\begin{equation}
A=\frac{r_{0}-r_{s}}{r_{0}^{3}}\textnormal{ and } \mu(r)=1-\frac{r_{s}}{r},
\end{equation}
the path equation becomes
\begin{equation}
\biggl(\frac{dr}{rd\phi}\biggr)^{2}=Ar^{2}-\mu.
\label{eq:Path2}
\end{equation}
Next, considering equation (\ref{sm}), setting $dl^2=0$, $\theta=\pi/2$ and dividing both sides by $dt^2$, we have
\begin{equation}
c^{2}\mu^{2}=\dot{r}^{2}+\mu(r\dot{\phi})^{2}
\label{eq:csqr}
\end{equation}
where $\dot{r}=v_{r}$ is the radial velocity component and $r\dot{\phi}=v_{\phi}$
is the tangential velocity component. Using the path equation (\ref{eq:Path2}),
we can solve for $v_{r}$ and $v_{\phi}$ in turn, to get:
\begin{equation}
v_{r}=\dot{r}=\pm c\mu\sqrt{1-\frac{\mu}{Ar^{2}}}
\label{eq:vr}
\end{equation}
\begin{equation}
v_{\phi}=r\dot{\phi}=\pm\frac{c\mu}{\sqrt{Ar^{2}}}
\label{eq:vphi}
\end{equation}
Thus the velocity vector of the photon along its path is 
\[
\boldsymbol{v}=v_{r}\boldsymbol{e_{r}}+v_{\phi}\boldsymbol{e_{\phi}}=\pm c\mu\left[\sqrt{1-\frac{\mu}{Ar^{2}}}\boldsymbol{e_{r}}\pm\frac{1}{\sqrt{Ar^{2}}}\boldsymbol{e_{\phi}}\right].
\]
To determine the acceleration vector, take the derivative with respect
to time:
\[
\boldsymbol{a}=\dot{v_{r}}\boldsymbol{e_{r}}+v_{r}\dot{\boldsymbol{e_{r}}}+\dot{v_{\phi}}\boldsymbol{e_{\phi}}+v_{\phi}\dot{\boldsymbol{e_{\phi}}}
\]
In polar coordinates, the derivatives of the unit vectors are $\dot{\boldsymbol{e_{r}}}=\dot{\phi}\boldsymbol{e_{\phi}}$ and $\dot{\boldsymbol{e_{\phi}}}=-\dot{\phi}\boldsymbol{e_{r}}$,
and so the acceleration components $a_{r}$ and $a_{\phi}$ are
\[
a_{r}=\dot{v_{r}}-\dot{\phi}v_{\phi},
\]
\[
a_{\phi}=\dot{v_{\phi}}+\dot{\phi}v_{r}.
\]
Differentiating $v_{r}$ in equation (\ref{eq:vr}) yields:
\[
\dot{v_{r}}=\pm\frac{c\dot{r}}{r^{2}}\left[r_{s}\sqrt{1-\frac{\mu}{Ar^{2}}}+\frac{\mu(2r-3r_{s})}{2Ar^2\sqrt{1-\mu/Ar^{2}}}\right].
\]
Equations (\ref{eq:vr}) and (\ref{eq:vphi}) are now used to eliminate the square root terms,
so that $\dot{v_{r}}$ simplifies to
\[
\dot{v_{r}}=\pm\frac{1}{r^{2}}\biggl[\frac{r_{s}v_{r}^{2}}{\mu}+(r-\frac{3}{2}r_{s})v_{\phi}^{2}\biggr]
\]
The radial acceleration component is therefore
\begin{equation}
a_{r}=\dot{v_{r}}-\dot{\phi}v_{\phi}=\dot{v_{r}}-\frac{v_{\phi}^{2}}{r}=\pm\frac{1}{r^{2}}\biggl[\frac{r_{s}v_{r}^{2}}{\mu}+(r-\frac{3}{2}r_{s})v_{\phi}^{2}\biggr]- \frac{v_{\phi}^{2}}{r}.
\label{vrdot}
\end{equation}
The acceleration must be related directly to the
Schwarzschild radius $r_{s}$ of the mass. There are two terms in
equation (\ref{vrdot}) that do not have an $r_{s}$ coefficient. These
two terms cancel if the positive sign is chosen. Thus, the correct form for the radial acceleration is:
\[
a_{r}=\frac{r_{s}}{r^{2}}\left[\frac{v_{r}^{2}}{\mu}-\frac{3v_{\phi}^{2}}{2}\right].
\]
A similar treatment for tangential acceleration component yields
\[
a_{\phi}=\dot{v_{\phi}}+\dot{\phi}v_{r}=\frac{r_{s}}{r^{2}}\frac{v_{r}v_{\phi}}{\mu}.
\]
Thus the acceleration vector for a photon near a Schwarzschild mass is:
\begin{equation}
\boldsymbol{a}=\frac{r_{s}}{r^{2}}\biggl{[}\bigl{[}\frac{v_{r}^{2}}{\mu}-\frac{3v_{\phi}^{2}}{2}\bigr{]}\boldsymbol{e_{r}}+\frac{v_{r}v_{\phi}}{\mu}\boldsymbol{e_{\phi}}\biggr{]}.
\label{aaa}
\end{equation}

It is interesting to note that the radius for light to remain in a circular orbit about the mass can immediately be derived from this acceleration vector. In a circular orbit, there is no radial velocity, and so
\begin{equation}
\boldsymbol{a}=\frac{r_{s}}{r^{2}}\left[-\frac{3v^{2}}{2}\right]\boldsymbol{e_{r}}.
\label{circ1}
\end{equation}
In addition, an object moving in a circular orbit in a classical kinematical framework has centripetal acceleration vector
\begin{equation}
\boldsymbol{a}=-\frac{v^{2}}{r}\boldsymbol{e_{r}}.
\label{circ2}
\end{equation}
When equations (\ref{circ1}) and (\ref{circ2}) are equated, we obtain
\begin{equation}
r=\frac{3r_{s}}{2}
\label{circ3}
\end{equation}
Thus, a photon at the `$3/2$' radius given in equation (\ref{circ3}) is trapped in a circular orbit about the mass, in accordance with the known result predicted by general relativity (\citet{car}, p 212).

\subsection{Kinematic Ray Shooting}

\begin{figure}
\vspace{1cm}
\caption{Path of photon from perihelion, plotted in Matlab using 2d kinematic ray shooting}
\includegraphics[width=7.5cm]{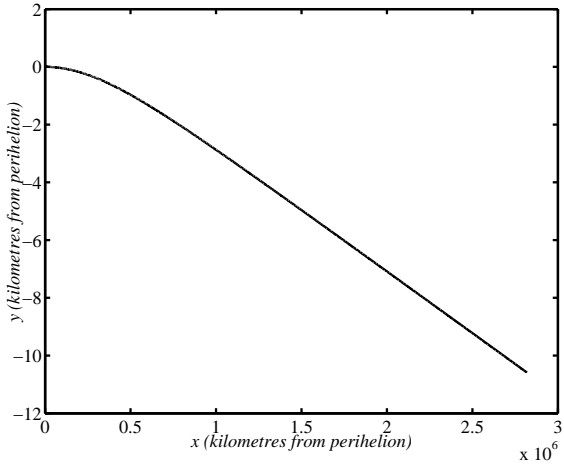}
\label{peri1}
\end{figure}

With known acceleration components, it is now possible to set up a
standard system of differential equations for ray tracing in polar
coordinates. The kinematical system is
\begin{eqnarray}
\frac{d}{dt}\left[\begin{array}{c}
r\\
\phi\\
v_{r}\\
v_{\phi}
\end{array}
\right]=\left[\begin{array}{c}
v_{r}\\
v_{\phi}/r\\
a_{r}+v_{\phi}^{2}/r\\
a_{\phi}-v_{r}v_{\phi}/r
\end{array}
\right]
\label{krt}
\end{eqnarray}
This new system (\ref{krt}) makes use of the new acceleration formula in equation (\ref{aaa}). Such a system can be solved using forward integration. Initial conditions for the photon have initial position at perihelion of 696000 kilometres
about a mass with Schwarzschild radius of 2.95 kilometres, zero radial
velocity, and tangential velocity $v_{\phi}=c/n=c\sqrt{\mu}$ to the
right. The speed of light is taken to be $c=300000$ kilometres/second. Using Matlab's ODE45 routine (an explicit Runge-Kutta 4-5 method) produces the path shown in Fig. (\ref{peri1}) for
the section shown; beyond this, the path is almost a straight line and so is not shown.
The slope of the line between the last integration point before the photon reaches earth orbit, and the first point after is $1.74851634\pm10^{-8}$ arcseconds. The time delay is calculated in the same integration, and is found to be $129.089609\pm10^{-6}$ $\mu$seconds. The uncertainty is due to limitations on the precision of Matlab's ODE45 routine. Both of these values correspond well with the predictions from general relativity (as calculated above), more closely than the first order approximations, and use of a higher precision
computation will allow a more accurate result, should such be required. Accuracy beyond first order is not commonly required, but these results give confidence that the acceleration vector presented here does accurately embody the effect on the photon due to a single Schwarzschild-type mass.

\begin{figure}
\vspace{1cm}
\caption{Path of photon past perihelion}
\includegraphics[width=7.5cm]{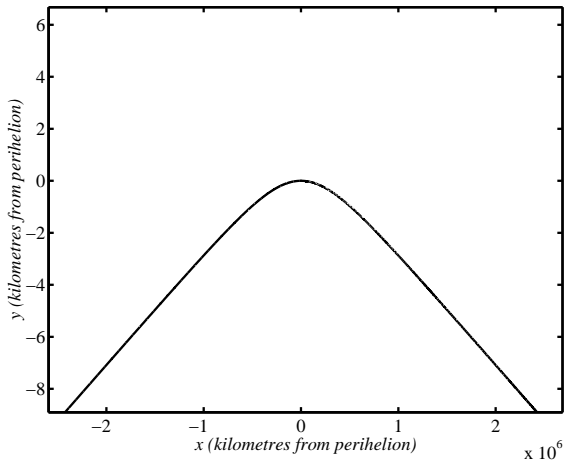}
\label{peri2}
\end{figure}

Using the values found for position and radial and tangential velocities
when the photon reaches earth orbit, we can send the photon along
the path from earth orbit past the sun and back out to earth orbit. The
central section is shown in Fig. (\ref{peri2}). Note that the scales differ by a factor of $10^6$.

The values for the deflection and the Shapiro delay calculated by this forward integration ray shooting are compared with the predictions from general relativity and the usual first order approximations(as calculated earlier) in Table \ref{t2}, and demonstrate that the kinematic ray shooting method presented in this paper is an accurate representation of the effect of the gravitational field of a single Schwarzschild body on the motion of a photon, giving us some confidence in using this method in approximating more complicated systems.
\begin{table}
\begin{tabular}{|c|c|c|}
\hline 
Approach & Deflection (arcsec) & Delay ($\mu$s)\tabularnewline
\hline
\hline 
Gen. Rel. prediction by integration along path (eqns. \ref{deflect1} and \ref{delay3}) & $1.74851634161261$ & $129.0896086$ \tabularnewline
\hline 
Usual approximations to first order in $r_{s}/r_{0}$ (eqns. \ref{deflect2} and \ref{eq:weinberg1})& $1.74850913341648$ & $129.0894053$
\tabularnewline
\hline 
Forward integration using new acceleration formula  & $1.74851634\pm10^{-8}$ & $129.089609\pm10^{-6}$\tabularnewline
\hline
\end{tabular}
\caption{A Comparison of the new method with accurate predictions from general relativity, and the common first order approximations.}
\label{t2}
\end{table}

\subsection{Magnification Maps}

When considering a multi-body system, such as a planetary system, it must be stressed that there is no known metric. That is, there is no known exact solution to Einstein's equations for such a system. Some sort of approximation is therefore required. Use of the `weak-field metric' is one such approximation, as is the addition of the deflection angles due to each body (the method generally used in microlensing models). Here, we choose to approximate by adding the acceleration components due to each body in the system.

Having tested the radial and tangential acceleration components described
above, it is a simple matter to set up a three dimensional ray tracing
system for a planetary system. At each integration point along each ray, the acceleration
components due to each massive body are calculated. This is done by a translation to put the massive body at the origin, followed by three rotations to place the photon's position vector and its velocity vector in the same plane as the massive body, with $\theta=\pi/2$. The radial and tangential velocities are then used to calculate the radial and tangential acceleration components. The three rotations are then reversed, and the resultant cartesian acceleration components are added to the acceleration components due to any other masses in the system to determine the overall acceleration of the photon.

\begin{figure}
\vspace{1cm}
\caption{Path of a photon through a supermassive binary black hole system. The black holes are indicated on the diagram with asterisks, and are located at (-946.08,-315.36) and (0,23.652).}
\includegraphics[width=16cm,height=8cm]{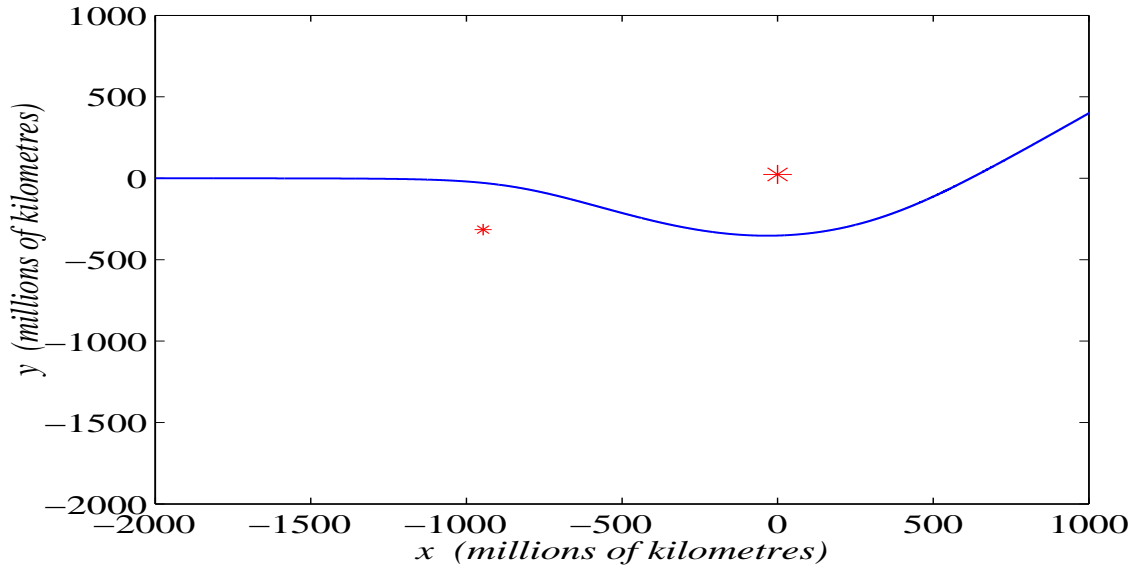}
\label{figure12}
\end{figure}

As a very simple (and artificial) example of this process, we first consider a two dimensional system, consisting of two very massive bodies in close proximity, and plot the path of the photon through this binary system. The smaller and larger stars represent bodies of 20 million and 50 million solar masses respectively (similar to the system described in \citet{bor}, although in the present example we imagine that the system has decayed to the point where the black holes are only a billion kilometres apart). This example is purely to demonstrate the versatility of the present approach. We are making the gross simplification that the black holes are stationary throughout the period when the ray is passing through the system, and so the acceleration of the two masses towards each other is therefore being ignored. For the purpose of demonstrating the procedure used here, we are ignoring such limitations. The present model is clearly a coarse approximation in this extreme case. The light ray comes in from the far left, and is deflected by the summed acceleration components due to each mass. The path is shown in Fig. (\ref{figure12}).

We now consider a more commonly modeled lensing system, the planetary lens. In describing a lensing system, it is common to use a parameter called the Einstein radius. This is the angular radius at which observers perfectly aligned with the point source and point lens would see a ring of light about the lens.
Specifically, the system is designed as follows: a point source is at $(-8000,0,0)$ and the observer
is at $(+8000,0,0)$. The lens star is placed at the origin, having Schwarzschild radius:
$r_{s}=99*10^{-8}$. An orbiting planet is placed at $(0,0.1208,0)$ (in microlensing terms, it is at $1.35$ times the Einstein radius), with $r_{s}=1*10^{-8}$. For simplicity in this model, we ignore the motion of the planet.
Rays are sent through the system, near the Einstein radius, and in
the vicinity of the planet. Due to symmetry in the cases here, it is only necessary to calculate half the rays and plot the result both above and below the axis of symmetry. During the numerical integration, each ray is broken into several small sections, with the size of each section becoming smaller as the photon nears the lens star. This is important to ensure that the integration routine does not take too large a step and miss the strong deflection altogether. For the simulations presented in this paper, each ray is broken into 36 segments. The result of this process is a light density map, or magnification
map (that is points in $(y,z)$ where the rays cross the plane $x=8000$). These simulations were run on MatLab version 6.1, under Windows XP, on an Athlon 64 3500+ processor with 1GB of ram. Running times are given as an indication, but no measures have been made to optimise the code for efficiency.

\begin{figure}
\vspace{1cm}
\caption{Caustic structure due to planet with mass 1\% of star's mass, located at 1.35 Einstein radii, using over 40,000 rays}
\includegraphics[width=16cm,height=8cm]{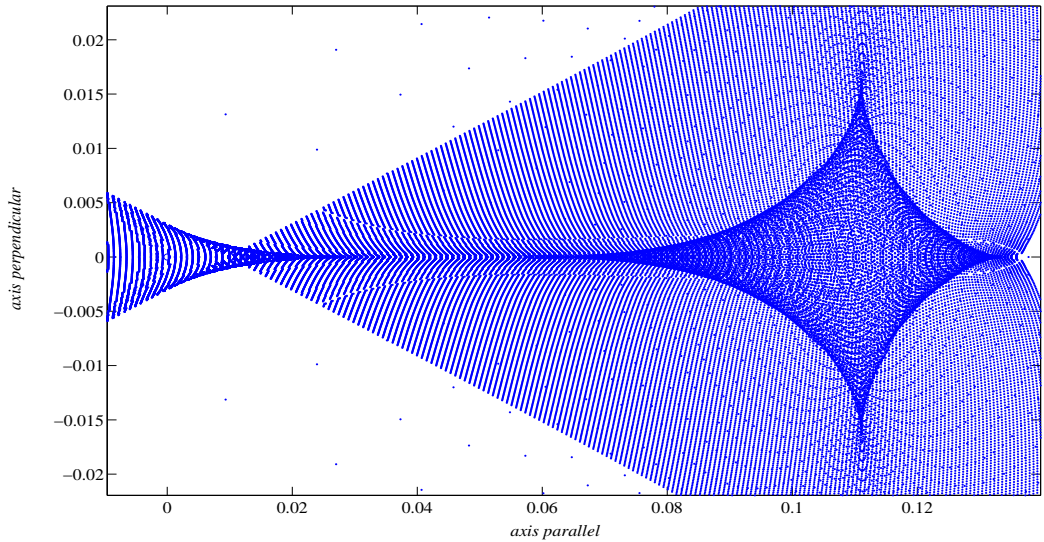}
\label{figure10}
\end{figure}

\begin{figure}
\vspace{1cm}
\caption{magnification density plot; same parameters as in Fig. (\ref{figure10}), using 106000 rays}
\includegraphics[width=16cm,height=8cm]{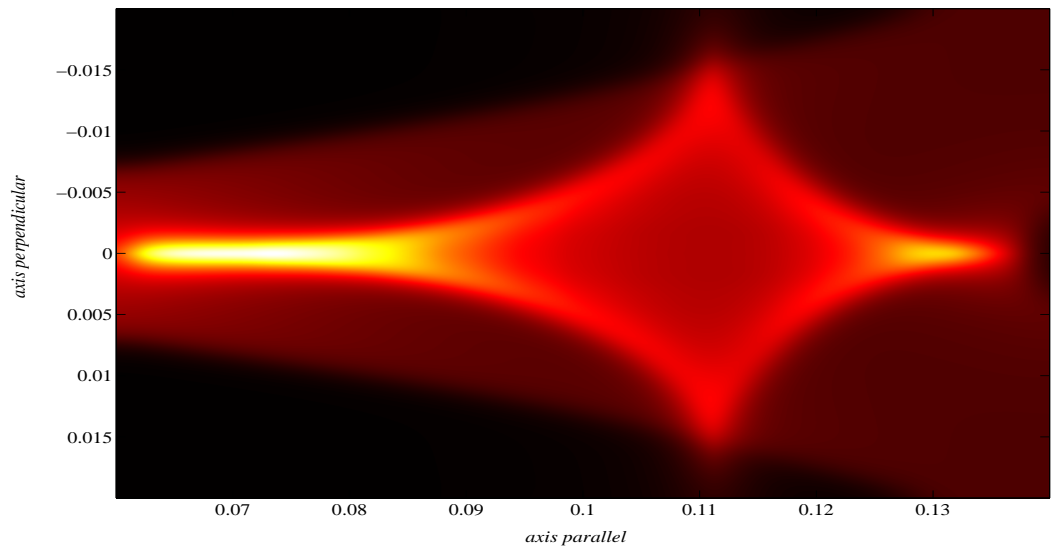}
\label{figure11}
\end{figure}

Fig. (\ref{figure10}) shows the magnification map produced when a rectangular array of rays (222 by 205) is sent through the lens system. The bending of the light towards the planet is clearly visible. This, combined with the bending caused by the lens star, produces the characteristic diamond shape for a system with a single planet outside the Einstein radius (\citet{wam}). The running time was 26 hours.
In Fig. (\ref{figure11}), many more rays have been used (approximately 106,000 rays). In order to view the resulting density, it is necessary to colour or shade regions according to how many rays pass through each small area. The code used to do this was `smoothhist2d', \citet{per}. The running time was approximately 50 hours. This result clearly shows the caustic diamond structure.

This diamond structure is the expected shape for the magnification map, and suggests that the method used here can be considered as an alternative method for modeling a thin lens.
\section{Conclusion and Discussion}

We have considered the path of a photon near a Schwarzschild-type body. Using the Schwarzschild metric, a new refractive index has been derived. Integrating the angle along the path gives the total deflection angle, and integrating the new index along the path gives the travel time delay. These values for the delay time and deflection can be calculated, using the formulae here, to arbitrary precision. This is because these formulae are derived directly from the Schwarzschild metric. As a check, it was shown that the standard first order approximations used for the deflection and delay agree with these results to first order.

A new formula for the acceleration of a photon was derived by combining the Schwarzschild metric with the path equation, and differentiating. This new acceleration formula was tested with a ray shooting approach, using the new refractive index to provide the initial conditions for the velocity components. The deflection and delay values were found to be in excellent agreement with the precise values calculated earlier.

By making the approximation that the acceleration on the photon is the sum of the individual acceleration components due to each massive body, a simple microlensing model was developed to demonstrate a use of the new acceleration formula for a binary system. Sample light fields on an observer plane have been computed using this new approach, and reproduce the expected figures. No attempt has been made here to speed computations, since that was not the purpose of the present work, but future developments may address such issues of computational efficiency.

In summary, this work provides a `classical' way of accurately describing the gravitational effect on a photon due to a single mass, and provides an alternative method for approximating the course of a photon through a complicated mass distribution. The authors believe that the approach presented here provides an insight into the effect of gravitating bodies on light rays that can be grasped without requiring a deep understanding of general relativity, and yet is still quantitatively accurate for a single mass, and can be used for approximating more complicated systems. As for the standard thin lens `deflection angle' method, this approach to gravitational lensing may be used by applied mathematicians, computer modellers and others without requiring specialist knowledge of general relativity. Because this approach retains the `delay' information as well as the deflection, it might conceivably be useful in analysing systems where the time delay plays a role, such as a pulsar source being lensed, should we observe such an event. It is also hoped that the formulae presented here will prove useful in producing models of more complicated mass distributions, such as galaxy clusters. Such models could be produced using the same method used here, simply by adding more bodies to the model.

\section*{Acknowledgement}
The authors are indebted to O. Wucknitz for constructive comments on an earlier draft of this paper.

\end{document}